\documentclass[%
 reprint,
superscriptaddress,
bibnotes,
 amsmath,amssymb,
 aps,
]{revtex4-2}

\usepackage[utf8]{inputenc}
\usepackage{siunitx}
\usepackage{array}
\usepackage{physics}
\usepackage{mathtools}
\usepackage{dsfont}
\usepackage[english]{babel}			
\usepackage{xcolor}
\usepackage{graphicx}				
\usepackage{amssymb,amsmath}		
\usepackage{url}					
\usepackage{marvosym}				
\usepackage[T1]{fontenc}            %
\usepackage{wrapfig}				
\usepackage{charter} 				
\usepackage{blindtext}				
\usepackage{datetime}				
\usepackage{lipsum}                 
\usepackage{float}                  
\usepackage{tabularx}
\usepackage{dcolumn}
\usepackage{bm}
\usepackage{todonotes}
\usepackage{hyperref}
\hypersetup{colorlinks=true,allcolors=blue}
\usepackage{enumitem}

\newcommand{\kettwo}[1]{$|#1\rangle$}

\newcommand{\ketbrathree}[1]{|#1\rangle\langle#1|}

\newcommand{\groundstateKET}{|\mathrm{G}\rangle}
\newcommand{\groundstateBRA}{\langle\mathrm{G}|}

\newcommand{\bareBrightHorizKET}{|\mathrm{X_H}\rangle}
\newcommand{\bareBrightHorizBRA}{\langle\mathrm{X_H}|}

\newcommand{\bareBrightVertiKET}{|\mathrm{X_V}\rangle}
\newcommand{\bareBrightVertiBRA}{\langle\mathrm{X_V}|}

\newcommand{\bareDarkHorizKET}{|\mathrm{D_H}\rangle}
\newcommand{\bareDarkHorizBRA}{\langle\mathrm{D_H}|}

\newcommand{\bareDarkVertiKET}{|\mathrm{D_V}\rangle}
\newcommand{\bareDarkVertiBRA}{\langle\mathrm{D_V}|}

\newcommand{\bareBrightHorVerKET}{|\mathrm{X_{H/V}}\rangle}
\newcommand{\bareDarkHorVerKET}{|\mathrm{D_{H/V}}\rangle}

\newcommand{\mixeBrightHorizKET}{|\mathrm{\mathcal{X}_H}\rangle}

\newcommand{\mixeBrightHoriz}{\mathrm{\mathcal{X}_H}}

\newcommand{\mixeBrightVertiKET}{|\mathrm{\mathcal{X}_V}\rangle}
\newcommand{\mixeBrightVerti}{\mathrm{\mathcal{X}_V}}

\newcommand{\mixeBrightHVKET}{|\mathrm{\mathcal{X}_{H/V}}\rangle}

\newcommand{\mixeDarkHorizKET}{|\mathrm{\mathcal{D}_H}\rangle}

\newcommand{\mixeDarkHoriz}{\mathrm{\mathcal{D}_H}}

\newcommand{\mixeDarkVertiKET}{|\mathrm{\mathcal{D}_V}\rangle}

\newcommand{\mixeDarkHVKET}{|\mathrm{\mathcal{D}_{H/V}}\rangle}

\newcommand{\biexcitonstateKET}{|\mathrm{XX}\rangle}
\newcommand{\biexcitonstateBRA}{\langle\mathrm{XX}|}

\newcommand{\basisangle}{\zeta^\mathcal{P}}
\newcommand{\collectionangle}{\zeta^\mathrm{C}}

\newcommand{\pulseI}{\mathcal{P}^\mathrm{i}}
\newcommand{\pulseW}{\mathcal{P}^\mathrm{s}}
\newcommand{\pulseR}{\mathcal{P}^\mathrm{r}}

\newcommand{\UIBK}{Institut f{\"u}r Experimentalphysik, Universit{\"a}t Innsbruck, 6020 Innsbruck, Austria}

\newcommand{\TUdo}{Condensed Matter Theory, Department of Physics, TU Dortmund, 44221 Dortmund, Germany}
\newcommand{\JKU}{Institute of Semiconductor and Solid State Physics, Johannes Kepler University Linz, 4040 Linz, Austria}

\newcommand{\Bayreuth}{Theoretische Physik III, Universit{\"a}t Bayreuth, 95440 Bayreuth, Germany}

\preprint{APS/123-QED}

\begin{document}

\title{Keeping the photon in the dark} 


\title{Keeping the photon in the dark: Enabling full quantum dot control by chirped pulses and magnetic fields}



\author{Florian Kappe}
\altaffiliation{These authors contributed equally}
\affiliation{\UIBK}
\author{René Schwarz}
\altaffiliation{These authors contributed equally}
\affiliation{\UIBK}
\author{Yusuf Karli}
\altaffiliation{These authors contributed equally}
\affiliation{\UIBK}
\author{Thomas Bracht}
\affiliation{\TUdo}
\author{Vollrath M. Axt}
\affiliation{\Bayreuth}
\author{Armando Rastelli}
\affiliation{\JKU}
\author{Vikas Remesh}
\affiliation{\UIBK} 
\author{Doris E. Reiter}
\affiliation{\TUdo}
\author{Gregor Weihs}
\affiliation{\UIBK}

\date{Date: \today \\ \phantom{XXX} E-mail: florian.kappe@uibk.ac.at}
\begin{abstract}
Because dark excitons in quantum dots are not directly optically accessible, so far they have not played a significant role in using quantum dots for photon generation. 
They possess significantly longer lifetimes than their brighter counterparts and hence offer enormous potential for photon storage or manipulation. 
In this work, we demonstrate an all-optical storage and retrieval of the spin-forbidden dark exciton in a quantum dot from the ground state employing chirped pulses and an in-plane magnetic field. Our experimental findings are in excellent agreement with theoretical predictions of the dynamics calculated using state-of-the-art product tensor methods. 
Our scheme enables an all-optical control of
dark states without relying on any preceding decays. This opens up a new dimension for optimal quantum control and time-bin entangled photon pair generation from quantum dots.
\end{abstract}
\maketitle
\section{Introduction} \label{sec:Intro}

As the establishment of a quantum network \cite{lu2021quantum} is rapidly advancing semiconductor quantum dots have emerged as a promising and versatile platform\cite{frick2023single,vajner_quantum_2022,akopian_entangled_2006,jayakumar_deterministic_2013}.
Their ability to generate high quality states of quantum light, e.g. single photons or correlated multi photon states \cite{carosini2023programmable}, make them a prime candidate in quantum technology. 
Single photons and entangled photon pairs are essential resources of optical quantum computing \cite{briegel2009measurement,flamini2018photonic,maring2024versatile}, secure communication via quantum key distribution (QKD) \cite{bennett1984proceedings,ekert1991quantum} and the distribution of quantum information in general \cite{sangouard2012single}.
    
The solid-state nature of quantum dots allows engineering the spectral properties via the growth process \cite{da2021gaas} or post-fabrication tuning methods \cite{kuklewicz2012electro,trotta2012universal,grim_scalable_2019,gerardot2007manipulating}. In addition to these, the interaction with lattice vibrations (phonons) \cite{bracht2022phonon,reindl2017phonon, reiter_phonon_2012,Reiter2019} delivers a challenging but rewarding quantum landscape.

Sophisticated excitation protocols \cite{luker2019review, bracht2021swing, karli2022super, sbresny2022stimulated, koong_multiplexed_2020, wilbur2022notch, kappe2024chirped} and the implementation into photonic structures \cite{heindel2023quantum} have put quantum dots at the 
forefront of quantum emitters producing single photons with record values in aspects like the single photon purity \cite{hanschke_quantum_2018}, indistinguishability \cite{Somaschi2016}, photon counts \cite{Tomm2021} and control over coherence properties \cite{karli2024controlling}.

The photon generation from quantum dots relies on the recombination of bright excitons, i.e. the radiative recombination of an electron / heavy hole pair of opposite spin. 
In addition, excitons consisting of parallelly oriented spins exist which are optically inactive \cite{bayer2000spectroscopic, poem2010accessing,zielinski2015atomistic, schwartz2016deterministic, luker2015direct, luker_phonon_2017,heindel2017accessing, jimenez2017dark, germanis2021emission, vargas2022dark,solovev2022manipulation}.
Due to suppression of emission, these so called dark exciton states exhibit a significantly reduced decay rate, resulting in lifetimes that can be orders of magnitude longer than their optically bright counterparts \cite{germanis2021emission,solovev2021long}. This attribute makes them ideal candidates for storing and distributing quantum coherence over time  \cite{schwartz2016deterministic}, e.g., in the generation of time-bin entanglement \cite{aumann_demonstration_2022,simon2005creating}.





To exploit the full potential of the dark exciton and enabling control over long-lived states and delayed photon emission, one has to address the challenge of optically accessing and coherently manipulating it. For this, one possibility is to prepare states in higher excited manifolds \cite{poem2010accessing, schwartz2015deterministic}, relying on subsequent decays. Preparing the dark exciton coherently within the ground state manifold has remained a theoretical proposal \cite{luker2015direct, luker_phonon_2017,neumann2021optical}.

In this work we propose and implement a simple, and flexible method to prepare and manipulate optically dark states in a quantum dot using chirped picosecond laser pulses and an external in-plane magnetic field (Voigt configuration). More insights on the preparation of the dark state are obtained by state-of-the-art numerical simulations. Our novel method unlocks control over these often overlooked states, extending the utility of quantum dots as a platform for quantum applications.

\section{Characterization of the dark exciton}
\begin{figure*}[hbt]
    \includegraphics[width = \linewidth]{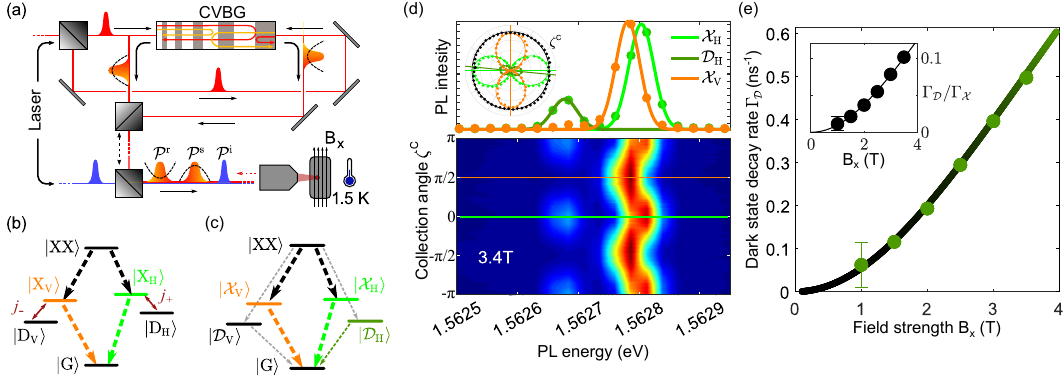}  
    \caption{\textbf{Dark state characterisation} 
    (a) Sketch of the system and experimental setup: Chirped laser pulses are prepared via a chirped volume Bragg grating (CVBG) and the help of 4f pulse shapers (not shown). Sequences up to three pulses called initialization ($\pulseI$, shown in blue), storage ($\pulseW$) and retrieval ($\pulseR$) pulse are sent onto a quantum dot hosted in a cryostat at \SI{1.5}{\kelvin} equipped with a vector magnet.
    (b) Level scheme of the quantum dot within the bare-state basis, consisting of the groundstate ($\groundstateKET$), the two bright exciton states ($\bareBrightHorVerKET$), the dark exciton states ($\bareDarkHorVerKET$) and the biexciton state ($\biexcitonstateKET$). Coupling induced by the magnetic field is abbreviated by $j_\pm$.  
    (C) Level scheme in the new eigenbasis including the magnetic field, displaying the mixed eigenstates $\mixeBrightHVKET$ and $\mixeDarkHVKET$. 
    Dashed downward arrows in (b) and (c) indicate optically active transitions and their coupling strength.
    (d) Results of the spectrally-resolved polarisation map from a magneto-photoluminescence (PL) measurement at \SI{3.4}{\tesla}. The collection angle $\collectionangle$ specifies the projection onto a linear polarisation state which is continuously varied in the range $[-\pi, \pi]$. Top panel: Line cuts at $\collectionangle = 0$ (green, $\mathcal{X}_\mathrm{H}$ and $\mathcal{D}_\mathrm{H}$) and $\collectionangle = \frac{\pi}{2}$ (orange, $\mathcal{X}_\mathrm{V}$) and corresponding Gaussian fits (solid lines) showing three distinct emission lines. Inset: Polarisation diagram for the three emission lines, extracted from the Gaussian fits. Black data points correspond to the total intensity of the bright states $\mixeBrightHorizKET$ and $\mixeBrightVertiKET$. Solid lines are fits to a sinusoidal oscillation in intensity. For clarity not all data points are included in the figure. (e) Decay rate measured from $\mixeDarkHorizKET$ ($\Gamma_\mathcal{D}$) for increasing magnetic field strength $\mathrm{B_x}$. The inset shows the degree of mixing defined by the ratio of decay rates from $\mixeDarkHorizKET$ ($\Gamma_\mathcal{D}$) and $\mixeBrightHorizKET$ ($\Gamma_\mathcal{X}$). Solid lines are best fits to the theoretical model.} 
    \label{figD1}
\end{figure*}

Direct optical manipulation of the dark state is enabled by a pulse sequence consisting of a transform-limited pulse and a pair of chirped pulses. All pulses are energetically tuned using 4\textit{f} pulse shapers and chirps of $\mp\SI{45}{\pico\second\squared}$ are introduced via reflection off a chirped volume Bragg grating \cite{kappe2023arp,kappe2024chirped}.
The quantum dot is hosted in a closed-cycle cryostat with a base temperature of \SI{1.5}{\kelvin} and optical activity of the dark state is enabled and controlled by a magnetic field of up to \SI{4}{\tesla} as indicated in Fig.~\ref{figD1}(a).


The electronic system of the quantum dot consists of the ground state $\groundstateKET$, the single exciton states and the biexciton state $\biexcitonstateKET$ as indicated in Fig.~\ref{figD1}(b). Without magnetic field, the energy eigenstates of the quantum dot can be divided into the bright states $\bareBrightHorVerKET$, where the spins of electron and hole are opposite and dark states $\bareDarkHorVerKET$ with parallelly oriented spins. Bright states can be excited by linearly polarised light in horizontal ($H$) or vertical ($V$) polarisation. With the same polarisation, these exciton states couple to the biexciton $\biexcitonstateKET$, i.e. the two photon emitting state \cite{huber2018strain,basset2023signatures}. The dark states in the simplest picture are optically inaccessible. That means, once prepared, they would not decay optically. In reality, valence band mixing and Coulomb mixing to higher excited exciton states could lead to brightening of these dark states \cite{poem2010accessing,zielinski2015atomistic,heindel2017accessing}. 

In our experiment, we induce a weak coupling between the bright and dark exciton states by a magnetic field $\mathrm{B_x}$ in Voigt configuration ($j_\pm \propto \mathrm{B_x} $ in Fig.~\ref{figD1} (a)). This leads to new eigenstates of the system, which we denote as $\mixeBrightHVKET$ and $\mixeDarkHVKET$ (see Sec. ~\ref{sec:App-new-states} for a detailed description). Now all states are optically coupled as indicated in Fig.~\ref{figD1}(c). Still, the coupling between the exciton states is quite different. For $\mixeBrightHVKET$ the optical coupling matrix element is strong, resulting in a fast decay, while for $\mixeDarkHVKET$ the optical coupling matrix element is rather weak with a slow decay rate. Hence, we keep the language from before and discriminate between bright $\mixeBrightHVKET$ and dark exciton $\mixeDarkHVKET$ states, even though the new eigenstates $\mixeDarkHVKET$ are not completely dark anymore. 

Initially, to identify the bright and dark states, a polarisation-resolved magneto-photoluminescence (PL) measurement is performed as shown in Fig.~\ref{figD1}(d). For optical excitation we use a \SI{635}{\nano\meter} continuous wave laser source, and set the magnetic field strength to \SI{3.4}{\tesla}.  Two bright emission lines around \SI{1.5628 }{\electronvolt} are identified, which vary out of phase as a function of collection polarisation angle $\collectionangle$. These lines correspond to emission from the transitions $\mixeBrightHorizKET \xrightarrow{} \groundstateKET$ and $\mixeBrightVertiKET \xrightarrow{} \groundstateKET$.

Besides the two bright exciton transitions, we observe an additional single dim emission line at \SI{1.56265}{\electronvolt}. We attribute this emission to stem from the $\mixeDarkHorizKET \xrightarrow{} \groundstateKET$ transition based on its energetic position and its oscillatory behaviour as function of the collection angle. 

To further confirm the assignment of the states, we compute the degree of linear polarisation $\mathrm{DOLP} = (I_{x'} - I_{y'}) / (I_{x'} + I_{y'})$. Here $I_{x'}$ is the maximum intensity and $I_{y'}$ is intensity along an orthogonal axis. We obtain for the three transitions the values $\mathrm{DOLP}_{\mixeBrightHoriz} = 0.92(1)$, $\mathrm{DOLP}_{\mixeBrightVerti} = 0.94(1)$ and $\mathrm{DOLP}_{\mixeDarkHoriz} = 0.73(1)$ which is in agreement with similar work \cite{germanis2018dark} andfurther corroborates our assumption of the identification of the dim line as a dark exciton. We note that the angle of maximum emission from $\mixeDarkHorizKET$ deviates from that of $\mixeBrightHorizKET$ by $\approx \SI{-0.1}{\radian}$, see Fig.~\ref{figD1} (d). 

The absence of emission from $\mixeDarkVertiKET$ suggests a highly anisotropic mixing behaviour between the two orientations of polarisation which makes $\mixeDarkVertiKET$ non-addressable for this specific quantum dot in its orientation in the vector magnet. A strong preference for mixing along one polarization can occur for certain combinations of parameters, i.e. Landé g-factors ($g_{ex}$ and $g_{hx}$), due to cancellations of the different parts (for an extended discussion see Section \ref{sec:theo_model}).

Using the information gained, the collection polarisation is aligned with the quantum dot horizontal (vertical) polarisation bases (light green, (orange) line in Fig.~\ref{figD1}(d)) by adjusting the collection angle $\collectionangle = 0(\frac{\pi}{2})$). Note that the terms horizontal and vertical just refer to the linear orthogonal polarisation states of the quantum dot. 

To underline the discrimination of bright and dark states, we measure the decay rates of both $\mixeBrightHorizKET$ and $\mixeDarkHorizKET$, denoted as $\Gamma_\mathcal{X}$ and $\Gamma_\mathcal{D}$ respectively, in Fig.~\ref{figD1} (e). As expected, the loss rate $\Gamma_\mathcal{D}$ increases for increasing magnetic field strength. The inset shows the ratio of $\Gamma_\mathcal{D}$/$\Gamma_\mathcal{X}$, reaching a maximum of $\approx 0.1$ for the fields achievable in our experimental setup. Thus, a strong difference in lifetimes remains between bright and dark states, justifying that we can still distinguish between two types of excitons and call them bright $\mixeBrightHVKET$ and dark $\mixeDarkHVKET$ states.


\section{Storage and retrieval utilizing the dark exciton}
\begin{figure*}[t]
    \includegraphics[width = \linewidth]{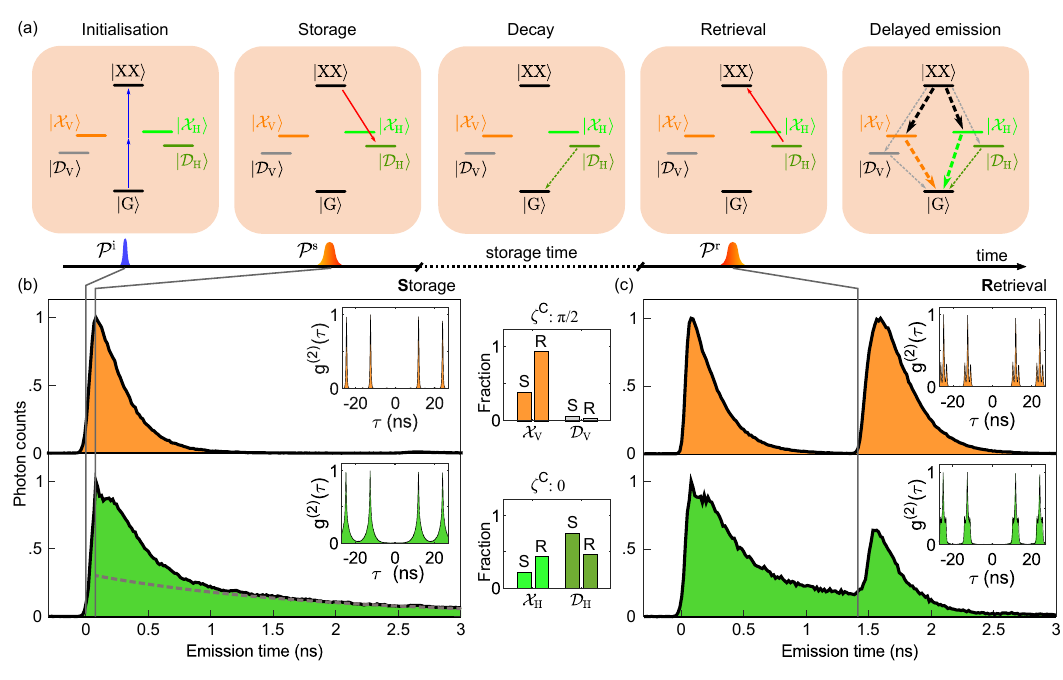}  
    \caption{
    \textbf{Time-resolved measurements and photon characteristics} (a) Sketches of the steps during the store-and-retrieve protocol 
    (b) Time-resolved photon emission during the storage sequence for vertical ($\collectionangle = \frac{\pi}{2}$, orange) and horizontal polarisation ($\collectionangle = 0$, green). A dashed grey line indicates the long time scale decay from $\mixeDarkHorizKET$. The insets show the second-order auto-correlation traces ($g^{(2)}(\tau)$).
    (c) The same as for (b) but including the retrieval pulse $\pulseR$. In the center panels we compare the integrated photon counts assigned to emission energies from $\mixeBrightHVKET$ and $\mixeDarkHVKET$ when switching from storage (S) to retrieval(R). }
    \label{figD2}
\end{figure*}

\subsection{Storage}
Having identified the dark exciton, the aim is to optically control its state in a storage and retrieval sequence of optical pulses as indicated in Fig.~\ref{figD2}(a). The full sequence comprises of a series of laser pulses with variable time delay between them. In this sequence, the storage process consists of the initialisation ($\pulseI$) and storage pulse ($\pulseW$). The first pulse $\pulseI$ brings the system from its ground to the biexciton state via two-photon excitation \cite{kappe2023arp,  karli2024controlling}. The following storage pulse $\pulseW$ prepares the system in the dark exciton state $\mixeDarkHorizKET$ starting from the biexciton. The preparation of the dark exciton is achieved by using a horizontally polarised, negatively chirped (GDD = \SI{-45}{\pico\second\squared}) laser pulse inducing an adiabatic evolution of states, similar to the suggestion in Ref.~\cite{luker2015direct}.

The storage sequence is monitored via the time-resolved photon emission under the application of the two pulses as shown in Fig.~\ref{figD2}(b). We note that all states are now optically active with differently strong dipole moments, such that radiative emission takes places at all times. We take two sets of data: (orange) Emission of $\mixeBrightVertiKET$ at $\collectionangle=\frac{\pi}{2}$ and $\mathrm{B_x} = \SI{2}{\tesla}$ and (green) emission from $\mixeBrightHorizKET$ and $\mixeDarkHorizKET$ at$\collectionangle=0$ and $\mathrm{B_x} = \SI{3.4}{\tesla}$. The action of the two pulses is marked by vertical lines in the figure.

Induced by the initialization pulse, the biexciton becomes occupied, resulting in a cascaded decay behaviour following $\biexcitonstateKET \xrightarrow{} \mixeBrightHVKET / \mixeDarkHVKET \xrightarrow{} \groundstateKET$ accompanied by an immediate rise in photon counts. 


At $ \approx \SI{0.07}{\nano\second}$ the storage pulse $\pulseW$ is applied. This interrupts the decay of the biexciton into the exciton states and accordingly the rise of the photon emission. Detecting the vertical polarisation (orange), after the storage pulse we observe only the emission $\mixeBrightVertiKET \xrightarrow{} \groundstateKET$ indicated by an abrupt transition to single exponential decay, signaling the depopulation of $\biexcitonstateKET$ induced by $\pulseW$.
While in an ideal storage process this emission would vanish, the decay during the initialisation pulse up to the storage pulse already leads to an occupation of $\mixeBrightVertiKET$. 

In the case of horizontal detection (green) the behaviour is different and with the storage pulse, a double exponential decay from $\mixeBrightHorizKET$ and $\mixeDarkHorizKET$ sets in. In addition, we find a sharp feature during the storage pulse $\pulseW$. Here a transient occupation during the pulse of the bright exciton $\mixeBrightHorizKET$ leads to a strong increase of photon counts during $\pulseW$. Imperfections in the preparation protocol lead to population remaining in both the bright exciton and the biexciton, yielding the shoulder following the sharp feature at $ \approx \SI{0.25}{\nano\second}$. 

Remembering that the dark exciton eigenstate $\mixeDarkHorizKET$ has a finite lifetime (cf. Fig.~\ref{figD1}(e)) and eventually decays, the slow exponential decay stems then from the dark exciton with a rate of $\Gamma_\mathcal{D} \approx \SI{0.56}{\per\nano\second}$ indicated as a dashed gray line. Comparing the two cases (orange and green), already signals that we have prepared the dark exciton $\mixeDarkHorizKET$.

\subsection{Retrieving}
After waiting for a storage time sufficiently long for the states $\biexcitonstateKET$ and $\mixeBrightHVKET$ to fully decay ($\approx \SI{1.3}{\nano\second} > 3*(\Gamma_\mathcal{X})^{-1}$), we apply a positively (GDD = \SI{45}{\pico\second\squared}) chirped laser pulse to retrieve the dark exciton population and bring the system back into the biexciton state \cite{falco}. From the biexciton state a cascaded emission into the ground state takes place and is recorded in the photon emission.

In Fig.~\ref{figD2}(c) we show the data recorded for the whole protocol. For both collection angles we see that by the storage pulse again photon emission is triggered. 

It is important to compare the two shapes of emission peaks at the storage and retrieval steps for vertical collection polarisation (orange). In the storage step, we find mostly a single exponential decay starting abruptly after $\pulseW$ arrives (see Fig.~\ref{figD2}(b)). In contrast, in the retrieval step there is a rise followed by a smooth transition to an exponential decay. This is also obvious in the widths of the two patterns. This behaviour is typical for a cascaded decay, where the exciton is fed by the biexction, while simultaneously decaying into the ground state.  

The area under the second peak is set by the storage sequence, which determines the amount of population to be stored. As such, the second peak can be adjusted by the time difference between $\pulseI$ and $\pulseW$ since the decay between the pulses is related to the storeable population. We observe that in the vertical case, the second emission peak is about the same height as the first emission peak, while in the horizontal case, the second emission peak is less pronounced. 

After the arrival of $\pulseR$, i.e. at timescales longer than $\SI{1.5}{\nano\second}$, the dynamics is mostly governed by the decay 
via the bright exciton states.
A negligible response at long-time scales, i.e., after $\SI{2.5}{\nano\second}$, hints towards the small but finite optical activity of the dark exciton decay channel for horizontal (green) polarisation.

\begin{figure*}[hbt!]
    \includegraphics[width = \linewidth]{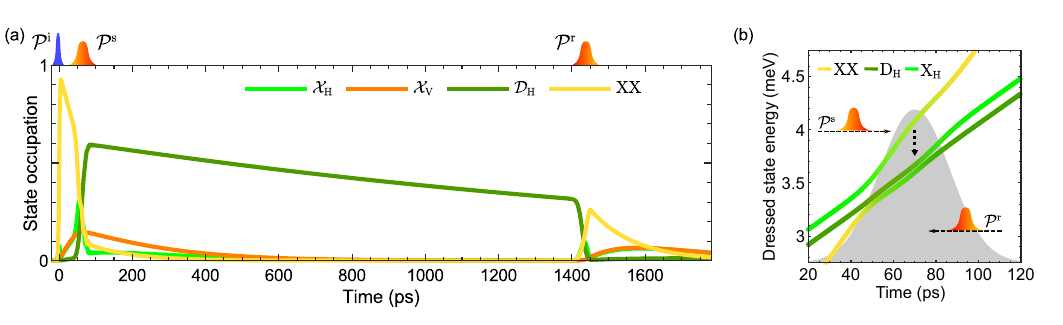}  
    \caption{\textbf{Simulation of the quantum dot dynamics}
    (a) Dynamics of the state occupations of the bright states $\mixeBrightHVKET$, the dark exciton state $\mixeDarkHorizKET$ and the biexciton state $\biexcitonstateKET$ under the pulse sequence displayed on top. 
    (b) Dressed state analysis of the system under the influence of $\pulseW$ ($\pulseR$ if read from right to left). The states $\bareBrightVertiKET$, $\bareDarkVertiKET$ and $\groundstateKET$ are effectively not influenced by $\pulseW$ ($\pulseR$) and are therefore excluded. Colors indicate the instantaneous eigenstate's overlap to the undisturbed bare states shown in Fig.~\ref{figD1}(b). The dotted downward arrow illustrates phonon relaxation processes.
    \label{figD3}
    }
\end{figure*}

\subsection{Photon counts and single photon character}
In the center of Fig.~\ref{figD2}, we quantify the amount of photons emitted during the storage (S) and retrieval (R) part by integrating the photon counts and discriminating them by their emission energy, indicated by $\mathcal{X}_{\mathrm{H/V}}$ and $\mathcal{D}_{\mathrm{H/V}}$. We note that only a negligible contribution from $\mixeDarkVertiKET$ is observed (a detailed explanation is shown in Sec.~\ref{sec:theo_model}). 

In the vertical (orange) case, the photon counts are normalized to the total photon emission during the full sequence, where we find that almost all photons are detected on the bright exciton line $\mathcal{X}_{\mathrm{V}}$. 
In the green case, i.e., for horizontal collection angle, the photon counts are normalized to the total emission during the storage sequence. Here, the emission originates mostly from the dark exciton $\mixeDarkHorizKET$.
Note that, the initial decay after $\pulseI$ but before $\pulseW$ arrives leads to still a finite emission from the bright exciton $\mixeBrightHorizKET$. After the full sequence the photon counts are almost equally distributed between the dark and bright exciton states. This is indicated by the green bar-plot in the center of Fig.~\ref{figD2}. By adjusting the parameters, this ratio could also be adjusted for a higher-efficiency preparation of $\mixeDarkHorizKET$. 

To characterise the nature of emission we also recorded the second-order auto-correlation traces ($g^{(2)}(\tau)$) for all four cases. We find that in all cases $g^{(2)}(0)$ is vanishing, proving the single-photon nature for both the recorded polarisations. In particular, this certifies that for the full sequence, the photon storage and retrieval has been successful.

\section{Theoretical analysis}

To understand the sequence to store and retrieve in and out of the dark state as well as the action of the chirped laser pulses, we performed numerical simulations of dynamics of the quantum dot states. For this, we set up a quantum dot model of the six electronic states and include the coupling to the external magnetic field as well as the pulsed optical driving. For the corresponding Hamiltonians we refer the reader to section \ref{sec:theo_model}. 

The full sequence comprises a series of laser pulses with the arrival times $t_0$ and the width $\tau$. The pulses have different frequencies $\omega_{{\cal{P}}}$ and can be of different polarisations $\mathbf{e}_{\mathcal{P}}$. Key to the protocol is allowing the storage and retrieval pulses to be chirped with the chirp coefficient $a$, such that the pulses read

\begin{equation}\label{equ:field}
    \begin{split}
        {\cal{P}} \sim& \mathbf{e}_{\mathcal{P}}\exp(-\frac{(t-t_0)^2}{2\tau^2}) \\
        &\times \exp(-i\left[\omega_\mathcal{P}+\frac{a(t-t_0)}{2}\right](t-t_0)).
    \end{split}
\end{equation}
More details on the driving fields are found in Sec.~\ref{sec:App_theory details}.

In addition we account for radiative decay by a Lindblad operator $\cal{L}$ via the rates $\Gamma_i$. We introduce the rates $\Gamma_\mathrm{X}$ and $\Gamma_\mathrm{XX}$ in the bare state system (see Fig.~\ref{figD1}(b)) that describe the decays of the bright states $\bareBrightHorVerKET \xrightarrow{} \groundstateKET$ and $\biexcitonstateKET\xrightarrow{} \bareBrightHorVerKET$, respectively, where the dark states do not couple to the light field and therefore do not decay. In the eigenstate picture (see Fig.~\ref{figD1}(c)) all transitions have a finite dipole moment and, accordingly, now all transitions are accompanied by a decay. Note that we perform all the numerical simulations in the bare state basis. 

We additionally account for the coupling to the phononic environment of the quantum dot. Phonons are known to be the major source of decoherence in quantum dot dynamics \cite{luker2019review,reiter_distinctive_2019}. We include the phonons on a microscopic level and solve the occurring many-body problem with a Process Tensor Matrix Product Operator (PT-MPO) method for a numerically exact description \cite{cygorek2022simulation,cygorek2024sublinear}. The simulation parameters are found in section \ref{sec:App_theory details} table \ref{tab:simulation}.

The results of the numerical calculations are shown in Fig.~\ref{figD3} (a). The occupations of the quantum dot states $\mixeBrightHVKET$, $\mixeDarkHorizKET$ and $\biexcitonstateKET$ are displayed under the action of the laser pulse sequence composed of $\pulseI$, $\pulseW$ and $\pulseR$ shown on top. Initially, only the ground state is occupied, such that all displayed occupations are zero. The first pulse $\pulseI$ leads to an occupation of the biexciton visible as strong rise of the biexciton state occupation. It is followed by the decay into the two bright excitons $\mixeBrightHVKET$.

The storage pulse $\pulseW$ brings most of the biexciton occupation into the dark exction state $\mixeDarkHorizKET$. During the pulse, we see a transient occupation of the horizontal bright exciton $\mixeBrightHorizKET$. This transient occupation results in the sharp feature observed in Fig.~\ref{figD2}(b). The other bright exciton $\mixeBrightVertiKET$ is unaffected by the pulse. 

After the storage pulse, a decay behaviour follows: On the one hand, there is still some biexciton occupation left, which decays mostly via the bright exciton states, which in turn decay into the ground state. The decay of the dark exciton state $\mixeDarkHorizKET$ occurs much slower. After half the storage time, i.e. around \SI{700}{\pico\second}, there is no occupation left in the bright states, while about 75\% of the dark exciton occupation after the pulse remains. After the full storage time we apply the retrieval pulse $\pulseR$, which switches the dark exciton occupation back to the biexciton state. From there, the cascaded decay ($\biexcitonstateKET \to \mixeBrightHVKET \to \groundstateKET$) takes place. 

To understand the action of the chirped pulses $\pulseW$ and $\pulseR$, which are key to our storage and retrival protocol, we consider their action in the dressed state picture \cite{bracht2023dressed}. The dressed state energies (i.e., the instantaneous eigenenergies of the coupled system) of the participating states in the rotating frame of $\pulseW$ are plotted in Fig.~\ref{figD3}(b). Colors represent the overlap of instantaneous eigenstates to the undisturbed bare states $\biexcitonstateKET$ (yellow), $\bareBrightHorizKET$ (light green) and $\bareDarkHorizKET$ (dark green).

During the action of the pulse an adiabatic evolution of states occurs. Because $\pulseR$ differs only in the sign of chirp from $\pulseW$, reading Fig.~\ref{figD3} (b) from right to left reveals the action of $\pulseR$. Via the adiabatic evolution during the pulses $\pulseW$ and $\pulseR$ the following states are connected:
\begin{itemize}
    \item $\bareBrightHorizKET \xrightleftharpoons[\pulseR]{\pulseW}\biexcitonstateKET$
    \item $\bareDarkHorizKET \xrightleftharpoons[\pulseR]{\pulseW} \bareBrightHorizKET$
    \item $\biexcitonstateKET \xrightleftharpoons[\pulseR]{\pulseW} \bareDarkHorizKET$
\end{itemize}
For our protocol, that means, we use both times the lowest dressed states going from $\biexcitonstateKET \to \mixeDarkHorizKET$ (yellow to dark green) with $\pulseW$ and reverse with $\pulseR$. Note that during the evolution the dressed state are mainly characterized by $\bareBrightHorizKET$ (light green segment in the middle), corresponding to the transient occupation in the dynamics. 

Phonons have little effect on the overall population dynamics, which is confirmed by simulations with and without phonons (See Sec.~\ref{sec:App_theory details}). This can be mainly deduced from the fact that the evolution along the lowest dressed state at low temperatures does not induce phonon transitions \cite{luker_influence_2012}.

\section{Performance analysis}
In reality, the successful preparation of $\mixeDarkHorizKET$ is dictated by the parameters of $\pulseW$ and limited by finite pulse durations, temporal separation of the pulses and more importantly, the decay from $\biexcitonstateKET$. 
\begin{figure*}[hbt!]
    \includegraphics[width = \linewidth]{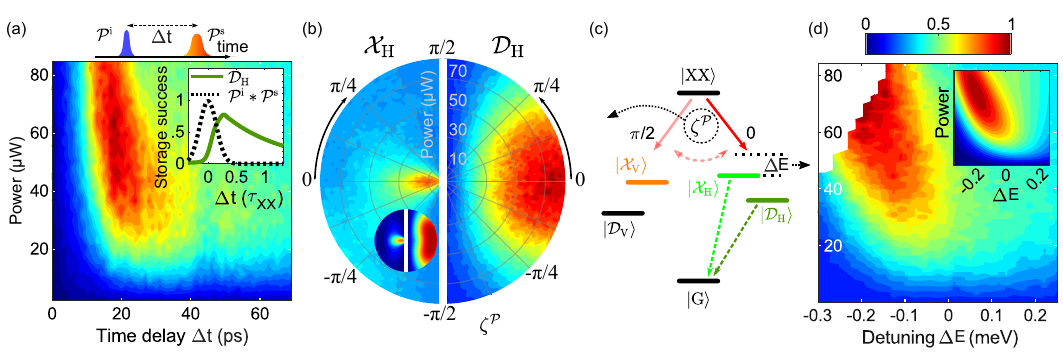}  
    \caption{\textbf{Dark state preparation characterisation} (a) Photon counts recorded from $\mixeDarkHorizKET$ for a variation of the temporal separation of $\pulseI$ and $\pulseW$ ($\Delta \mathrm{t}$) and the pulse power of $\pulseW$. The inset shows a numerical calculation of the storage fidelity at optimum power in units of the lifetime of $\biexcitonstateKET$ ($\tau_\mathrm{XX}$) alongside the normalised convolution of $\pulseI$ and $\pulseW$ ($\pulseI * \pulseW$). (b) Photon emission from $\mixeBrightHorizKET$ (left) and $\mixeDarkHorizKET$ (right) for varying angle of linear polarisation ($\basisangle$) of $\pulseW$. Pulse power is varied along the radial axis and for symmetry reasons only data in the range $\basisangle = [-\frac{\pi}{2},\frac{\pi}{2}]$ is shown. The inset shows a numerical calculation of the idealised process excluding decays. (c) Illustration of the parameters investigated in (b) and (d). (d) The same as for (a) but for varying detuning of $\pulseW$ ($\Delta \mathrm{E}$) relative to the emission energy $\biexcitonstateKET \xrightarrow{} \mixeBrightHorizKET$. The inset shows the numerical calculation of the idealised process, excluding decays.  Photon counts shown in (a), (b) and (d) are normalised to their respective maximum. All measurements were performed at a magnetic field strength of $\mathrm{B_x} = \SI{3.4}{\tesla}$.}
    \label{fig:Dark_prep}
\end{figure*}

Therefore we investigate the preparation efficiency of $\mixeDarkHorizKET$ by studying three parameters of the storage pulse $\pulseW$: 
\begin{enumerate}
    \item \textit{Time delay} between $\pulseI$ and $\pulseW$: $\Delta \mathrm{t}$
    \item \textit{Polarisation} of $\pulseW$: $\basisangle$  
    \item Energetic \textit{detuning} of $\pulseW$ with respect to the transition $\biexcitonstateKET \xrightarrow{} \mixeBrightHorizKET$: $\Delta \mathrm{E}$   
\end{enumerate}
In Fig.~\ref{fig:Dark_prep} we summarise the results. The efficiency of the dark state preparation is monitored by the integrated photon counts at the corresponding emission energy (see Fig.~\ref{figD1} (d)). All measurements are performed at a magnetic field strength of $\mathrm{B_x} = \SI{3.4}{\tesla}$.  

\textbf{Time delay:}
The storage step relies on the application of a storage pulse $\pulseW$ after initial preparation of $\biexcitonstateKET$ via a two photon resonant initialisation pulse $\pulseI$. The storage pulse $\pulseW$ needs to be negatively chirped to induce the adiabatic passage from $\biexcitonstateKET$ to $\mixeDarkHorizKET$. As a consequence, the transition into $\mixeDarkHorizKET$ cannot happen instantaneously, but requires a finite pulse length (cf. Eq.~\eqref{equ:field} in Sec.~\ref{sec:App_theory details}). Consequently, this imposes a finite time delay $\Delta \mathrm{t}$ between $\pulseI$ and $\pulseW$, during which the decay from $\biexcitonstateKET$ after $\pulseI$ lowers the preparation fidelity of $\mixeDarkHorizKET$. 

In Fig.~\ref{fig:Dark_prep} (a), we present a two-dimensional map of the measured $\mixeDarkHorizKET$ population as a function of the time delay $\Delta \mathrm{t}$ between $\pulseI$ and $\pulseW$, and power of $\pulseW$. For all powers of $\pulseW$, a maximum population of $\mixeDarkHorizKET$ is found at $\Delta \mathrm{t} \approx \SI{20}{\pico \second}$.

To understand the limits imposed by a finite $\biexcitonstateKET$ lifetime ($\tau_\mathrm{XX}$) and pulse durations, we simulate the storage success given by the occupation of $\mixeDarkHorizKET$. For this, we set the $\pulseW$ power to its optimal value and and study the dependence on $\Delta \mathrm{t}$. We calculate the temporal overlap of $\pulseI$ and $\pulseR$ as the normalised convolution of both field amplitudes ($\pulseI * \pulseW$) and present the results in units of $\tau_\mathrm{XX}$. The inset in Fig.~\ref{fig:Dark_prep} (a) shows that efficient population of $\mixeDarkHorizKET$ can only be achieved when $\Delta \mathrm{t}$ is sufficiently large such that $\pulseI * \pulseW \ll 1$, meaning the pulses should not overlap substantially. At larger $\Delta \mathrm{t}$ the fidelity is then further reduced by a finite $\tau_\mathrm{XX}$.

\textbf{Polarisation:}
In Fig.~\ref{fig:Dark_prep} (b) we study the preparation efficiency of $\mixeDarkHorizKET$ depending on the angle of linear polarisation $\basisangle$ and optical power of $\pulseW$ (as sketched in Fig.~\ref{fig:Dark_prep}(c)). For low powers the pulse sequence $\pulseI + \pulseW$ leads to an increase of emission from $\mixeBrightHorizKET$ compared to pure two-photon resonant excitation (power of $\pulseW$ = 0, inner-most data points), if the polarisation is aligned parallel to the transition $\biexcitonstateKET \xrightarrow{} \mixeBrightHorizKET$ ($\basisangle = 0$). In the case of orthogonal polarisation ($\basisangle = \pm \frac{\pi}{2}$) a suppression of emission is observed, see left side of Fig.~\ref{fig:Dark_prep} (b). This response has been observed before, e.g. in Ref.~\cite{karli2024controlling}. While this process shows a strong dependence on $\basisangle$ we find that the preparation of $\mixeDarkHorizKET$ is less sensitive to $\basisangle$ and happens at higher powers (right side of Fig.~\ref{fig:Dark_prep} (b)). We attribute this lowered sensitivity to the anisotropic mixing of $\bareBrightHorizKET \leftrightarrow \bareDarkHorizKET$ and $\bareBrightVertiKET \leftrightarrow \bareDarkVertiKET$, effectively favouring a transition into $\mixeDarkHorizKET$ at higher pulse powers. 

\textbf{Detuning:}
In Fig.~\ref{fig:Dark_prep} (d) we present the recorded photon emission from $\mixeDarkHorizKET$ when the central energy of $\pulseW$ is tuned with respect to the transition energy $\biexcitonstateKET \xrightarrow{} \mixeBrightHorizKET$ ($\Delta \mathrm{E}$, see Fig.~\ref{fig:Dark_prep}(c)) and its power is varied. We compare the recorded emission to numerical simulations of $\mixeDarkHorizKET$ population presented in the inset and find an optimum value for $\Delta \mathrm{E}$ of about \SI{-0.2}{\milli\electronvolt}, or $\approx \SI{1.5590}{\electronvolt}$ absolute energy in both cases. Because of the robustness of the adiabatic passage, a small detuning of the central frequency maintains a high final dark state population \cite{schmidgall2010population,kappe2023arp}. This feature can also be beneficial in a multi-level system \cite{luker_phonon_2017}.

\section{Conclusions} \label{sec:Discussion}
In summary, we have designed and demonstrated a novel method of storing and retrieving population utilizing the dark state in a semiconductor quantum dot. We have also provided an in-depth understanding of our quantum emitter system with theoretical simulations. With the usage of the external magnetic field and the chirped laser pulses, the system stays within the ground state manifold of the quantum dot and the optical control processes are coherent. As our protocol relies on pulses detuned from the exciton energy, it offers the advantage of simple spectral filtering being efficient for laser light suppression.
More importantly, using the dark exciton could improve several protocols to generate time-bin entanglement states \cite{simon2005creating, jayakumar_time-bin_2014, prilmuller2018hyperentanglement, aumann_demonstration_2022} or photonic cluster states \cite{istrati2020sequential,cogan2023deterministic}. Our results help leverage the versatile quantum dot state manifold including the optically dark exciton states and further expand the possibility of using quantum dots for quantum communication.

\section{Materials and methods} \label{sec:Methods}

\subsection{Experimental methods} \label{sec:MethodsExp}

The quantum dot is hosted in a closed-cycle helium cryostat (ICEOxford) with a base temperature of about \SI{1.5}{\kelvin} and a superconducting vector magnet system (up to \SI{4}{\tesla} absolute value).
We excite the quantum dot and collect photons from the top through a cold objective (numerical aperture 0.81, attocube systems AG), employing a cross-polarisation measurement setup for scattered laser light suppression. The collection polarisation basis is chosen linear and can be rotated by means of a half-wave plate (HWP).  Collected photons are sent through a home-built monochromator based on narrowband notch filters (BNF-805-OD3, FWHM \SI{0.3}{\nano\meter}, Optigrate), which is set up to collect single-photon emission and spectrally block scattered laser reflection. Measurements on spectral composition were performed using a single-photon sensitive spectrometer (Acton SP-2750, Roper Scientific) equipped with a liquid nitrogen cooled charge-coupled device camera (Spec10 CCD, Princeton Instruments). Time-sensitive measurements were carried out on superconductiong nanowire single photon detectors (Eos, Single Quantum) connected to a time-tagging module (Time Tagger Ultra, Swabian Instruments). The overall time-jitter of the single-photon measurement apparatus was \SI{20}{\pico\second}. 

\begin{figure*}[hb!]
\centering
\includegraphics[width = 1\linewidth]{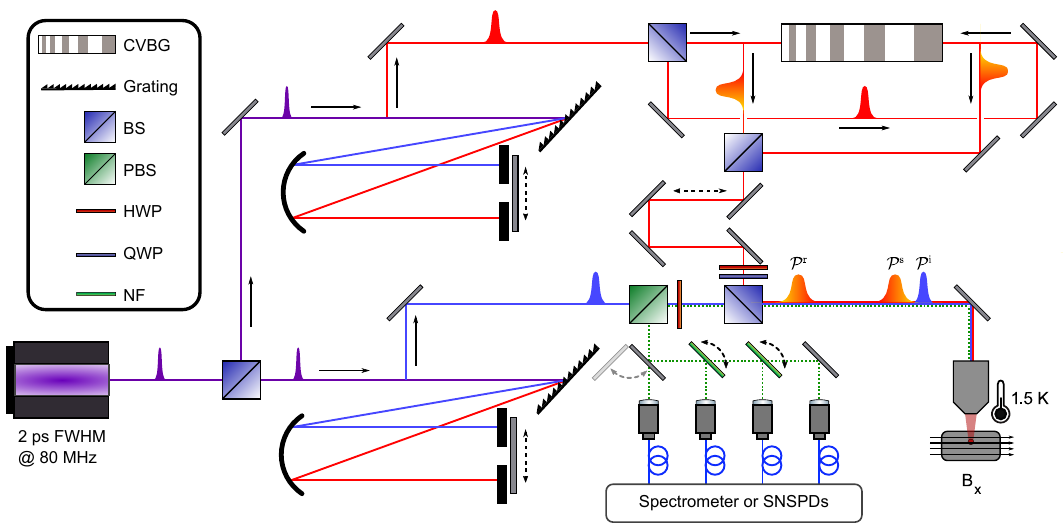}
\caption{\textbf{Schematic sketch of the experimental setup} For the pulse preparation a pulsed Ti:Sa laser with a pulse duration of $\approx \SI{2}{ps}$ FWHM, a repetion rate of $\approx \SI{80}{\mega\hertz}$ and a mean wavelength at $\SI{795}{nm}$ is split via a beam-splitter (BS) into two paths to spectrally shape pulses $\pulseI$, $\pulseW$ and $\pulseR$. Using folded $4$f pulse shapers consisting of a grating, a concave mirror as well as an motorized slit for spectral tunability. After shaping, laser light is split again and guided into a Mach-Zehnder like setup, leading to a chirped volume Bragg grating (CVBG). The reflected pulses are picking up a spectral chirp of $\mathrm{GDD} = \pm \SI{45}{ps^2}$ depending on the direction of reflection. The path of the $\SI{45}{ps^2}$ pulse ($\pulseR$) is designed to delay the pulse by $\approx \SI{1.3}{ns}$ with respect to the negatively chirped pulse ($\pulseW$). In each arm variable neutral density-filters are used for individual power control (not shown). After recombination, an additional variable time delay is introduces to adjust the relative arrival time between $\pulseI$ and $\pulseW$. Laser pulses are then coupled into the QD experiment setup, employing a cross-polarisation setup consisting of a polarising beam-splitter (PBS), half-wave plates (HWP) as well as quater-wave plates (QWP) for additional polarisation control ($\basisangle$) and collection basis ($\collectionangle$) selection. All pulses are recombined on another BS and directed into the cryostat with a base temperature of $\SI{1.5}{K}$ in which our quantum dot is placed. The quantum dot is under the influence of a in-plane magnetic field $\mathrm{B_x}$. Finally, the emitted photons from the quantum dot are coupled into a monochromator, consisting of reflective notch-filters (NF) on a rotation mount for frequency selection and pump light suppression, leading to either superconducting nano-wire single photon detectors (SNSPDs) or a spectrometer.}
\label{fig:setup}
\end{figure*}

The sequence of excitation pulses is created by sending a single $\approx\SI{2}{\pico \second}$ full width at half maximum (FWHM) long pulse (Tsunami 3950, SpectraPhysics) through two individual 4-f pulseshapers with variable slit positions. We prepare one pulse to be TPE resonant with a spectral bandwidth of $\approx \SI{0.2}{\nano\meter}$ FWHM. The other pulse is spectrally centred around the biexciton emission wavelength and also of $\approx \SI{0.2}{\nano\meter}$ FWHM. 
We split this pulse at a 50:50 beam splitter (BS) and send both onto a chirped volume bragg grating (CVBG, Optigrate). This prepares both pulses to carry $\SI{\mp45}{\pico\second\squared}$ of spectral chirp, depending of the direction of reflection. We delay the positively chirped pulse by $\approx \SI{1.3}{\nano\second}$ before recombining both on a second BS. The set of chirped pulses are delayed with respect to the TPE resonant pulse by a  fiber-optic delay line (ODL-300, OZ Optics). Recombination of both excitation paths happens at a 90:10 beamsplitter close to the cryostat entrance window. The polarisation of the chirped pulses can be arbitrarily chosen via the means of a HWP/QWP combination while the polarisation of the TPE pulse is fixed to be orthogonal to the collection polarisation via a polarising beam splitter (PBS). A detailed drawing of the experimental setup can be found in Fig.~\ref{fig:setup}.      

\subsection{Theoretical model} \label{sec:theo_model}
The Hamiltonian used to model the system dynamics can be written in the dipole- and rotating wave approximation, similar to our previous work \cite{karli2024controlling,kappe2023arp} as: 

\begin{equation} \label{equ:H_total}
    \hat{H} = \hat{H}^{\text{system-fields}} +\hat{H}^{\text{phonon}}.
\end{equation}

where $\hat{H}^{\text{system-fields}}$ treats the quantum dot system and its interaction with electromagnetic fields and reads as: 
\begin{equation}  \label{equ:H_light_matter}
    \hat{H}^{\text{system-fields}} =  \hat{H}^{\text{QD}} + \hat{H}^{\text{Bx}} + \hat{H}^{\text{L}}. 
\end{equation}
    
The individual parts are described in the following. 
We model the quantum dot as a six-level system in the linear polarisation basis consisting of the groundstate $\groundstateKET$, two bright excitons of horizontal ($\bareBrightHorizKET$) and vertical ($\bareBrightVertiKET$) polarisation, neighboring dark excitons ($\bareDarkHorizKET$ and $\bareDarkVertiKET$) and the biexciton state $\biexcitonstateKET$ (see Fig.~\ref{figD1} (b)). 

The groundstate energy is set to zero, while the two bright excitons are separated by a fine-structure splitting $\delta_X$, such that $E_\mathrm{X}^\mathrm{{H/V}} = \hbar\omega_\mathrm{X} \pm \frac{\delta_\mathrm{X}}{2}$. The biexciton has a binding energy $E_\mathrm{B}$ such that $E_\mathrm{{XX}} = 2\hbar\omega_\mathrm{X} -E_\mathrm{B}$. The dark states are treated similar to the bright states and posses a fine structure splitting $\delta_\mathrm{D}$, such that $E_\mathrm{D}^\mathrm{{H/V}} = \hbar\omega_\mathrm{D} \pm \frac{\delta_\mathrm{D}}{2}$. The quantum dot hamiltonian then reads

\begin{equation} \label{equ:H_QD}
    \begin{split}
        \hat{H}^{\text{QD}} = & E_\mathrm{X}^\mathrm{H} \bareBrightHorizKET\bareBrightHorizBRA + E_\mathrm{X}^\mathrm{V} \bareBrightVertiKET\bareBrightVertiBRA\\ & + E_\mathrm{D}^\mathrm{{H}} \bareDarkHorizKET\bareDarkHorizBRA + E_\mathrm{D}^\mathrm{{V}} \bareDarkVertiKET\bareDarkVertiBRA \\ & + E_\mathrm{{XX}} \biexcitonstateKET\biexcitonstateBRA.
    \end{split}
\end{equation}

Coupling to an in-plane magnetic field in Voigt configuration is included in $\hat{H}^{\text{Bx}}$ and introduces a mixing between bright states and their dark counterparts via 

\begin{equation}\label{equ:H_Bx}
    \begin{split}
        \hat{H}^{\text{Bx}} = & -\mu_B \frac{\mathrm{B_x}}{2} (g_{hx} + g_{ex}) \bareBrightHorizKET\bareDarkHorizBRA  \\ & -\mu_B \frac{B_x}{2} (g_{hx} - g_{ex}) \bareBrightVertiKET\bareDarkVertiBRA + \text{h.c.} \\ 
    \end{split}
\end{equation}

Here, $g_{ex}(g_{hx})$ denote the Landé g-factors and the coupling of the electron(hole) to the magnetic field, while $\mu_B$ is the Bohr magneton. For better readability the coupling terms $-\mu_B \frac{\mathrm{B_x}}{2} (g_{hx} \pm g_{ex})$ are abbreviated by $j_{\pm}$ in Fig.~\ref{figD1} (c). The asymmetry of $j_+$ and $j_-$ allows for strong anisotropic mixing if $|g_{ex}| \approx |g_{hx}|$ as is the case for the quantum dot used in this work, i.e. $j_+ \gg j_-$.     

The laser driving fields are coupled to the quantum dot via 

\begin{equation} \label{equ:H_tpe}
    \begin{split}
        \hat{H}^{L} = & \Omega(\Delta \mathrm{t},\omega_\mathcal{P},\Theta,\mathrm{GDD},\mathbf{e_L}) \cdot \\ (&\mathbf{e_H} (\bareBrightHorizKET\groundstateBRA+ \biexcitonstateKET\bareBrightHorizBRA) + \\ & \mathbf{e_V} (\bareBrightVertiKET\groundstateBRA + \biexcitonstateKET\bareBrightVertiBRA) + \text{h.c.}).
    \end{split}    
\end{equation}

Here $\mathbf{e_H}$ and $\mathbf{e_V}$ are vectors of unit length and represent the vectorial overlap of the horizontal and vertical dipole-moments to the laser polarisation ($\mathbf{e_L}$) via the dot product (see also Fig.~\ref{figD1} (d)).

The coupling term $\Omega(\Delta \mathrm{t},\omega_\mathcal{P},\Theta,\mathrm{GDD},\mathbf{e_L})$ is treated as classical chirped laser fields of Gaussian shape, see Equ.~\ref{equ:field} and Sec.~\ref{sec:App_theory details}.

We consider radiative decay and losses by a Lindblad operator $\cal{L}$. For a density matrix $\hat{\rho}$, a rate $\Gamma$ and operators $\hat{\cal{O}}$, the Lindblad operator takes the form

\begin{equation}
    {\cal{L}}_{\Gamma, \hat{\cal{O}}} [\hat{\rho}] = \Gamma \left(\hat{\cal{O}} \hat{\rho} \hat{\cal{O}}^{\dagger} - \frac{1}{2} \left\{\hat{\rho}, \hat{\cal{O}}^\dagger \hat{\cal{O}}\right\} \right),
\end{equation}

with $\{.,.\}$ indicating the anticommutator.
We incorporate radiative decays such that the Lindblad-superoperators read:
\begin{equation}
\begin{split}
    \label{equ:LindbladDecay}
    {\cal{L}}[\hat{\rho}] := & {\cal{L}}_{\Gamma_\mathrm{X},\groundstateKET\bareBrightHorizBRA}[\hat{\rho}]  + {\cal{L}}_{\Gamma_\mathrm{X},\groundstateKET\bareBrightVertiBRA}[\hat{\rho}] \\   
    & +{\cal{L}}_{\Gamma_\mathrm{XX},\bareBrightHorizKET\biexcitonstateBRA}[\hat{\rho}]  + {\cal{L}}_{\Gamma_\mathrm{XX},\bareBrightVertiKET\biexcitonstateBRA}[\hat{\rho}].
\end{split}
\end{equation}
The rates $\Gamma_\mathrm{X}$ and $\Gamma_\mathrm{XX}$ describe the decays $\bareBrightHorizKET  /\bareBrightVertiKET \xrightarrow{} \groundstateKET$ and $\biexcitonstateKET\xrightarrow{} \bareBrightHorizKET / \bareBrightVertiKET$ respectively and are indicated by downward dashed arrows in Fig.~\ref{figD1} (b). Note that for the bare states no decay from the dark states is contained in the model ($\Gamma_\mathrm{D} = 0$).

Only the inclusion of $\hat{H}^{\text{Bx}}$ leads to a mixing of bright and dark states and the formation of new eigenstates in the time-independent Hamiltonian 
$\hat{H}^{\text{QD}}+\hat{H}^{\text{Bx}}$. We therefore introduce the new states $\mixeBrightHorizKET$, $\mixeDarkHorizKET$, $\mixeBrightVertiKET$ and $\mixeDarkVertiKET$ as the eigenstates of $\hat{H}^{\text{QD}}+\hat{H}^{\text{Bx}}$. These states inherit properties of their respective parent states, including the coupling to the driving fields and radiative decay via photon emission, resulting in partially bright dark states as depicted in Fig.~\ref{figD1} (c). A decomposition of these states in terms of the original bare states and the new eigenenergies can be found in section \ref{sec:App-new-states}. 

Additionally we model dissipation via the coupling to longitudinal acoustic phonons by using the deformation potential coupling. With $\hat{b}_\mathbf{k}$ ($\hat{b}_\mathbf{k}^\dagger$) as annihilation (creation) operator of a phonon mode $\textbf{k}$ with frequency $\omega_\mathbf{k}$, the phonon coupling Hamiltonian reads as

\begin{equation}
    \hat{H}^{\mathrm{phonon}} = \hbar \sum_\mathbf{k} \omega_\mathbf{k} \hat{b}_\mathbf{k} \hat{b}_\mathbf{k}^\dagger + \hbar \sum_{\mathbf{k}, S} \left(\gamma_\mathbf{k}^S \hat{b}_\mathbf{k}^\dagger + {\gamma_\mathbf{k}^S}^\ast \hat{b}_\mathbf{k}\right) \ketbrathree{S}.
\end{equation}
Here each phonon mode $\mathbf{k}$ couples with a coupling constant $\gamma_\mathbf{k}^S$ to a quantum dot state \kettwo{S}, where $S \in \{\mathrm{X_H}, \mathrm{X_V}, \mathrm{D_H}, \mathrm{D_V}, \mathrm{XX}\}$. The coupling constant $\gamma_\mathbf{k}^S$ with its including material parameter are chosen to be the same as in Ref. \cite{Barth2016} and our previous works \cite{karli2024controlling,kappe2023arp}.

\subsection{Quantum dot sample} \label{sec:QD_sample}

The sample used contains GaAs/AlGaAs quantum dots obtained by the Al-droplet etching method \cite{da2021gaasS} and was grown by molecular beam epitaxy. The quantum dots are embedded in the center of a $\lambda$-cavity placed between a bottom(top) distributed Bragg reflector consisting of 9(2) pairs of $\lambda/4$-thick $\mathrm{Al}_{0.95}\mathrm{Ga}_{0.05}\mathrm{As}/ \mathrm{Al}_{0.20}\mathrm{Ga}_{0.80}\mathrm{As} $ layers with respective thicknesses of 69/60 nm. The quantum dots are placed between two $ \lambda/2$-thick $\mathrm{Al}_{0.33}\mathrm{Ga}_{0.67}\mathrm{As}$ layers. The quantum dot growth process starts by depositing 0.5 equivalent monolayers of Al in the absence of arsenic flux, which results in the self-assembled formation of droplets. During exposure to a reduced As flux, such droplets locally etch the underlying $\mathrm{Al}_{0.33}\mathrm{Ga}_{0.67}\mathrm{As}$ layer, resulting in $\approx \SI{9}{\nano\meter}$-deep and $\approx$ \SI{60}{\nano\meter} wide nanoholes on the surface. Then the nanoholes are filled with GaAs by depositing $\approx\SI{1.1}{\nano\meter}$ of $\mathrm{GaAs}$ on the surface, followed by an annealing step of \SI{45}{\second}. The temperature used for the etching of the nanoholes was \SI{600}{\celsius}. The droplet self-assembly process results in quantum dots with random position and a surface density of about $2\times 10^{7} \si{\per\centi\meter\squared}$, suitable for single quantum dot spectroscopy. We note that the same sample was also used in our previous works \cite{kappe2023arp,karli2024controlling}.

\section{Acknowledgments and declarations} \label{sec:Acc_and_Dec}
FK, RS, YK, VR, and GW acknowledge the financial support through the Austrian Science Fund FWF projects with grant IDs 10.55776/TAI556 (DarkEneT), 10.55776/W1259 (DK-ALM Atoms, Light, and Molecules), 10.55776/FG5, 10.55776/I4380 (AEQuDot) and FFG. For open access purposes, the authors have applied a CC-BY public copyright license to any author-accepted manuscript version arising from this submission. AR acknowledges the FWF projects 10.55776/FG5, 10.55776/P30459, 10.55776/I4320, the Linz Institute of Technology (LIT), the European Union's Horizon 2020 research, and innovation program under Grant Agreement Nos. 899814 (Qurope), 871130 (ASCENT+), and the QuantERA II Program (project QD-E-QKD, FFG Grant No. 891366). TKB and DER acknowledge financial support from the German Research Foundation DFG through project 428026575 (AEQuDot). 
\bibliography{ref.bib}%

\clearpage

\section{Appendix} 

\subsection{Quantum dot states in Voigt configuration} \label{sec:App-new-states}
As described in the main text we utilize an in-plane magnetic field in Voigt configuration to introduce a mixing of the initial bare bright states $\bareBrightHorizKET$ and $\bareBrightVertiKET$ with the initial dark states $\bareDarkHorizKET$ and $\bareBrightVertiKET$. 
This leads to the formation of new eigenstates ($\mixeBrightHorizKET,~\mixeBrightVertiKET, ~ \mixeDarkHorizKET ~ \text{and} ~ \mixeDarkVertiKET$) and new eigenergies via the diagonalisation of the time independent Hamiltonian $\hat{H}^{\text{QD}}+\hat{H}^{\text{Bx}}$. Since $\hat{H}^{\text{Bx}}$ does not introduce any mixing on them and for better readability the states $\groundstateKET$ and $\biexcitonstateKET$ are omitted and we restrict our self to the four dimensional subspace spanned by single excitonic states: 
\begin{equation*}
    \begin{pmatrix}
    E_\mathrm{X}^\mathrm{H} & 0 & j_{+} & 0 \\
    0 & E_\mathrm{X}^\mathrm{V} & 0 & j_{-} \\
    j_{+} & 0 & E_\mathrm{D}^\mathrm{H} & 0 \\
    0 & j_{-} & 0 & E_\mathrm{D}^\mathrm{V} \\
    \end{pmatrix}
    \xrightarrow{}
    \begin{pmatrix}
    
     E_\mathcal{X}^\mathrm{H} & 0 & 0 & 0\\
    0 & E_\mathcal{X}^\mathrm{V} & 0 & 0\\
    0 & 0 & E_\mathcal{D}^\mathrm{H} & 0\\
    0 & 0 & 0 & E_\mathcal{D}^\mathrm{V}\\
    
    \end{pmatrix}
\end{equation*}

Here $j_\pm = -\mu_B \frac{\mathrm{B_x}}{2} (g_{hx} \pm g_{ex})$ introduces the magnetic field dependent mixing. The new eigenenergies read as: 
\begin{equation*}
    \begin{split}
        E_\mathcal{X}^\mathrm{H} = & \frac{1}{2}(+\sqrt{(E_\mathrm{X}^\mathrm{H}-E_\mathrm{D}^\mathrm{H})^2 + 4 j_+^2} + E_\mathrm{X}^\mathrm{H} + E_\mathrm{D}^\mathrm{H}) \\
        E_\mathcal{X}^\mathrm{V} = & \frac{1}{2}(+\sqrt{(E_\mathrm{X}^\mathrm{V}-E_\mathrm{D}^\mathrm{V})^2 + 4 j_-^2} + E_\mathrm{X}^\mathrm{V} + E_\mathrm{D}^\mathrm{V})\\
        E_\mathcal{D}^\mathrm{H} = & \frac{1}{2}(-\sqrt{(E_\mathrm{X}^\mathrm{H}-E_\mathrm{D}^\mathrm{H})^2 + 4 j_+^2} + E_\mathrm{X}^\mathrm{H} + E_\mathrm{D}^\mathrm{H})\\
        E_\mathcal{D}^\mathrm{V} = & \frac{1}{2}(-\sqrt{(E_\mathrm{X}^\mathrm{V}-E_\mathrm{D}^\mathrm{V})^2 + 4 j_-^2} + E_\mathrm{X}^\mathrm{V} + E_\mathrm{D}^\mathrm{V}).\\
    \end{split}
\end{equation*}

Decomposing the new mixed states in terms of the bare states reads: 
\begin{equation*}
    \begin{split}
        \mixeBrightHorizKET = & \frac{1}{N}(\frac{E_\mathrm{X}^\mathrm{H}-E_\mathrm{D}^\mathrm{H}+\sqrt{(E_\mathrm{X}^\mathrm{H}-E_\mathrm{D}^\mathrm{H})^2+4j_+^2}}{2 j_+}\bareBrightHorizKET + \bareDarkHorizKET)\\
        \mixeBrightVertiKET = & \frac{1}{N}(\frac{E_\mathrm{X}^\mathrm{V}-E_\mathrm{D}^\mathrm{V}+\sqrt{(E_\mathrm{X}^\mathrm{V}-E_\mathrm{D}^\mathrm{V})^2+4j_-^2}}{2 j_-}\bareBrightVertiKET + \bareDarkVertiKET)\\ 
        \mixeDarkHorizKET = & \frac{1}{N}(\frac{E_\mathrm{X}^\mathrm{H}-E_\mathrm{D}^\mathrm{H}-\sqrt{(E_\mathrm{X}^\mathrm{H}-E_\mathrm{D}^\mathrm{H})^2+4j_+^2}}{2 j_+}\bareBrightHorizKET + \bareDarkHorizKET)\\ 
        \mixeDarkVertiKET = & \frac{1}{N}(\frac{E_\mathrm{X}^\mathrm{V}-E_\mathrm{D}^\mathrm{V}-\sqrt{(E_\mathrm{X}^\mathrm{V}-E_\mathrm{D}^\mathrm{V})^2+4j_-^2}}{2 j_-}\bareBrightVertiKET + \bareDarkVertiKET)\\ 
    \end{split}
\end{equation*}
with $N$ normalising the states to unit length. 

\subsection{Details on simulation} \label{sec:App_theory details}

As mentioned in the main text the driving fields acting on the quantum dot are treated as classical laser fields of Gaussian shape given by 

\begin{equation}\label{equ:field}
    \begin{split}
        \Omega(\Delta \mathrm{t},\omega_\mathcal{P},\Theta,\alpha,\mathbf{e_L}) = & \frac{\mathbf{e_L}\Theta}{\sqrt{2\pi\tau_0\tau}}\exp(-\frac{(t-\Delta \mathrm{t})^2}{2\tau^2}) \\ & \exp(-i(\omega_\mathcal{P}+\frac{a(t-\Delta \mathrm{t})}{2})(t-\Delta \mathrm{t})) 
    \end{split}
\end{equation}
with $t$ being the time and $\Delta \mathrm{t}$ the time of arrival, $\Theta$ being the pulse area, $\omega_\mathcal{P}$ the central frequency of the laser field and 
\begin{equation}
    \begin{split}
        \tau = & \sqrt{\frac{\text{GDD}^2}{\tau_0^2}+\tau_0^2}, \\
        a = & \frac{\text{GDD}}{\text{GDD}^2 + \tau_0^4}
    \end{split}
\end{equation}

incorporating the effect of spectral chirp $\text{GDD}$ onto the pulse shape in time with a transform limited duration of $\tau_0$.

The vector $\mathbf{e_L}$ describes the orientation of linear polarisation of the laser pulse and with respect to the polarisation basis of the quantum dot ($\mathbf{e_H}$ and $\mathbf{e_V}$) is given as

\begin{equation}
\label{equ:laser_polarisation}
    \mathbf{e_L} = \cos{\basisangle} \mathbf{e_H} + \sin{\basisangle} \mathbf{e_V}.
\end{equation}

Here $\basisangle$ denotes the angle with respect to emission from $\mixeBrightHorizKET$ (horizontal green line in Fig.~\ref{figD1} (d)).   

A summary of the used simulation parameters is presented in the following table:

\begin{table}[ht]
    \centering
    \caption{Simulation parameters. Values are either measured or estimated to our best knowledge from an assemble of quantum dots (QD) on the same sample.}
    \begin{tabular}{c|c c c c }
        Symbol & QD & $\pulseI$ & $\pulseW$ & $\pulseR$ \\
        \hline
       $\hbar \omega_\mathrm{X}$  & \SI{1.5628}{\electronvolt} & - & - & -\\
       $\hbar \omega_\mathrm{D}$  & \SI{1.5627}{\electronvolt} & - & - & -\\
       $E_\mathrm{B}$ &\SI{3.6}{\milli\electronvolt}& - &- &- \\
       $\delta_\mathrm{X}$ & \SI{11.14}{\micro\electronvolt}& -& -& -\\
       $\delta_\mathrm{D}$ & \SI{11.14}{\micro\electronvolt}& -& -& -\\
       $(g_{hx} + g_{ex})$ &0.41 & - & -& - \\
       $(g_{hx} - g_{ex})$ &0 & - & -& - \\
       $\Gamma_\mathrm{X} ^{-1}$ &\SI{180}{\pico\second} &- &- &- \\
       $\Gamma_\mathrm{XX} ^{-1}$ &\SI{120}{\pico\second} &- &- &- \\
       QD size & \SI{5}{\nano\meter} & - &- &- \\
       T & \SI{1.5}{\kelvin} & -&- &- \\
       $\mathrm{B_x}$ & \SI{3.4}{\tesla} & - & - & - \\
       $\hbar\omega_\mathcal{P}$ & - & \SI{1.5610}{\electronvolt}& \SI{1.5590}{\electronvolt}& \SI{1.5590}{\electronvolt} \\
       $\tau_0$ & - & \SI{2.9}{\pico\second} & \SI{2.9}{\pico\second} & \SI{2.9}{\pico\second} \\ 
       $\text{GDD}$ & - & \SI{0}{\pico\second\squared} & \SI{-45}{\pico\second\squared} & \SI{+45}{\pico\second\squared} \\
       $\basisangle$ & - & \SI{0}{\radian} & \SI{0}{\radian} & \SI{0}{\radian} \\
       $\Theta$ & - & 4.5 $\pi$& 3.5 $\pi$ & 3.5 $\pi$ \\
       $\Delta \mathrm{t}$ & - & \SI{0}{\nano\second} & \SI{0.07}{\nano\second} & \SI{1.42}{\nano\second}\\
          
    \end{tabular}
    
    \label{tab:simulation}
\end{table}

\subsubsection{Details on state dynamics}
\begin{figure}
    \centering
    \includegraphics{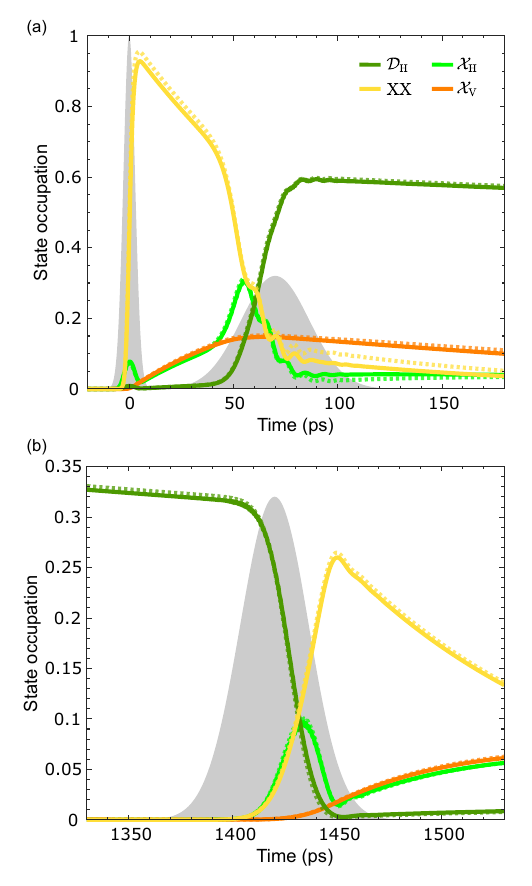}
    \caption{\textbf{State dynamics during storage and retrieval}
    (a) State dynamics during the initialisation pulse $\pulseI$ (at \SI{0}{\pico\second}) and the storage pulse $\pulseW$ (at \SI{70}{\pico\second}). Solid lines are simulations including phonon interaction while dashed lines exclude these interactions. (b) Same as in (a) but for the retrieval pulse $\pulseR$ at \SI{1420}{\pico\second}.}
    \label{fig:store_retrieve_zoom}
\end{figure}

In this section we pay closer attention to the quantum dot state dynamics during the storage and retrieval steps of the protocol, shown in Fig.~\ref{fig:store_retrieve_zoom}(a) and (b) respectively.

Assuming that the chirped pulse $\pulseW$ induces a nearly adiabatic system evolution along dressed states, see Fig.~\ref{figD3}(b), the simulation suggests that the initial $\biexcitonstateKET$ component (yellow) is transferred first to $\bareBrightHorizKET$ (light green segment in the middle) and eventually to the dark state $\bareDarkHorizKET$ (dark green) while the smaller initial $\bareBrightHorizKET$ component should end up in $\biexcitonstateKET$. This is reflected in Fig.~\ref{fig:store_retrieve_zoom}(a). The dynamic is dominated by the lowest dressed state which initially corresponds to a large $\biexcitonstateKET$ component. The adiabatic evolution then continuously decreases the $\biexcitonstateKET$ population. The intermediary transfer to $\bareBrightHorizKET$ is clearly seen as an intermediate maximum in the $\mixeBrightHorizKET$ population. The transfer of the initial $\bareBrightHorizKET$ population to $\biexcitonstateKET$ has little impact since it was rather small at the start but is likely responsible for the finding that the $\biexcitonstateKET$ population stays above that of $\mixeBrightHorizKET$. 

Phonons induce transitions between the dressed states. The dotted arrow in Fig.~\ref{figD3}(b) illustrates a phonon emission that slightly reduces the population of the upper dressed state, which at long times approaches $\biexcitonstateKET$. Thus, we expect a reduction of the $\biexcitonstateKET$ population which is indeed seen in Fig.~\ref{fig:store_retrieve_zoom}(a) where dotted lines stem from simulations without phonons while solid lines are calculated including phonons. Interestingly, the dark state population after the pulse $\pulseW$ is reduced by phonons although the dark state is the lowest dressed state at long times. We attribute this on the one hand to a phonon induced population exchange between the lower two dressed states during the pulse where these states are rather close in energy such that deviations from the adiabatic picture are likely and on the other hand to phonon absorption processes that can lead to a redistribution from a lower to a higher dressed state in particular when theses states are close in energy. This is in line with the observation that phonons increase the $\mixeBrightHorizKET$ population which would also not be expected if the dynamics were to adiabatically follow the dressed states. 

Reading Fig.~\ref{figD3}(b) from right to left reveals that a time inverted (positively chirped) pulse $\pulseR$ reverts population from $\bareDarkHorizKET$ back into $\biexcitonstateKET$. If applied sufficiently long after $\pulseW$ ($\Gamma_\mathcal{D}^{-1} > $ storage time $ \gg \Gamma_\mathcal{X}^{-1} > \Gamma_\mathrm{XX}^{-1}$) the only occupied state, besides $\groundstateKET$, is $\mixeDarkHorizKET$ which leads to an almost perfect population inversion from $\mixeDarkHorizKET \xrightarrow{} \biexcitonstateKET$, limited only by decay during the pulse. This process is shown in Fig.~\ref{fig:store_retrieve_zoom}(b) and highlights the population retrieval evoked by $\pulseR$ which transfers the dark state occupation back into bright states.

\subsection{Estimation of g-factors} \label{app:g-factors}

\begin{figure}[h!]
    \centering.
    \includegraphics[width = 1\linewidth]{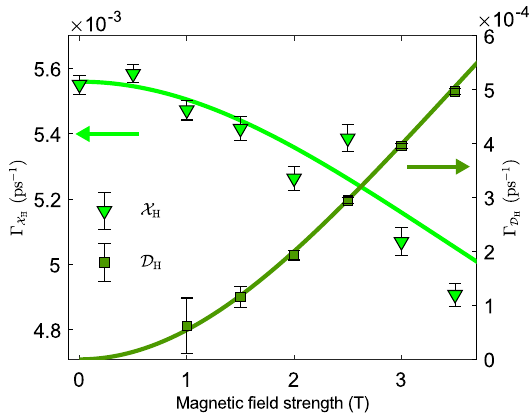}
    \caption{\textbf{Decay rate dependence} Magnetic field dependence of decay rates from $\mixeBrightHorizKET$ and $\mixeDarkHorizKET$. Solid lines are best fits to our numerical model resulting in a combined Landé g-factor $(g_{hx}+g_{ex}) = 0.411(3)$.}
    \label{fig:g_factor_fit}
\end{figure}

We estimate the combined g-factors $(g_{hx}+g_{ex})$ by stimulating from $\biexcitonstateKET$ into an equal superposition of $\mixeBrightHorizKET$ and $\mixeDarkHorizKET$ and recording time-resolved photon emission. We then extract the decay times of the participating rates by fitting a bi-exponential decay to the data. Repeating this for a set of different magnetic fields and fitting to our numerical model lets us estimate the combined Landé g-factors (see Fig.~\ref{fig:g_factor_fit}):

\begin{equation*}
    (g_{hx}+g_{ex}) = 0.411(3)  
\end{equation*}

Together with the observation $(g_{hx}+g_{ex}) \gg (g_{hx}-g_{ex})$ we estimate the individual g-factors as: 

\begin{equation*}
    g_{hx} = g_{ex} \approx 0.205.
\end{equation*}

\newpage
\subsection{Experimental setup} \label{sec:Setup}

\end{document}